
\magnification\magstep1
\openup 2\jot
\centerline{\bf Quantization of Bogomol'nyi soliton dynamics}
\vskip 0.3in
\centerline{M. Temple-Raston}
\vskip 0.2in
\centerline {Department of Mathematics,}
\vskip 0.1in
\centerline{Concordia University, 1455 de Maisonneuve Blvd.,}
\vskip 0.1in
\centerline {Montr\'eal, Qu\'ebec H4B 1R6 Canada}
\vskip 0.5in
\openup -2\jot
\bigskip
\noindent
{\bf Abstract.}
\medskip
\noindent
We approximate analytically the semi-classical differential cross-section
for low-energy solitonic BPS $SU(2)$ magnetic monopoles using the
geodesic approximation.  The semi-classical scattering amplitude,
$f(\theta)$, can be expressed as a conditionally convergent infinite
series which is made absolutely convergent by analytic continuation.
Our results suggest that the classical solitonic cross-section (computed
numerically elsewhere) and the semi-classical cross-sections are in good
agreement over a wide range of scattering angles,
$\pi/3<\theta<\pi/2$ and $\pi/2<\theta<2\pi/3$.

\openup 2\jot
\medskip
\noindent
{\bf Introduction.}
\medskip
Recent numerical studies on the interactive dynamics of two solitonic
BPS $SU(2)$ magnetic monopoles indicate that some characteristic
behaviour associated with the quantum dynamics of point-particles can be
observed classically when the particles are solitonic [1].
Examples of this `pseudo-quantum' behaviour include, one, that
the numerically generated low-energy classical bosonic differential
cross-section at $\pi/2$ scattering angle in the centre-of-mass frame
is in unexpected agreement with the quantum differential cross-section
given by a partial wave analysis, and, two, the classical stationary
bound states appear to have a discrete energy spectrum.
In this letter we extend the agreement in the classical solitonic and
quantum differential cross-sections to angles round $\theta=\pi/2$.
While the computations in this paper refer specifically to the
magnetic monopole, there is evidence to suggest that vortices [2],
Skyrmions [3], solitons in nonlinear $\sigma$-models [4], and other
solitons arising from a Bogomol'nyi structure have a similar classical
scattering behaviour to that of solitonic magnetic monopoles [5].
The mathematical structures associated with these other solitons are
weaker than that for the magnetic monopole, leading to fewer analytic
results.  As a result, more effort and sophisticated numerical work is
needed to compensate for the weaker structure.  The theoretical
development for magnetic monopoles should comment usefully
on these other solitons, particularly on those in three dimensions.

\smallskip
The underlying classical dynamics of solitonic magnetic monopoles at
low-energies is approximated by the geodesic approximation, for
details see [6,7].  The highly non-trivial classical dynamics
of the solitonic BPS $SU(2)$ magnetic monopole two-body problem
has been studied in detail [7-10].  Furthermore, Gibbons and Manton
have shown that by taking the monopole Schr\"odinger operator to be
proportional to the covariant Laplacian defined using the natural metric
on the moduli space of BPS magnetic monopoles, the correspondence principle
is respected [10].  While the Schr\"odinger operator proposed by
Gibbons and Manton is indeed very natural, our approach will be to
obtain a semi-classical theory which comments on all possible quantizations
without committing ourselves to a particular Schr\"odinger operator.
The classical dynamics are sufficiently well-understood analytically
and numerically to give a good approximation to the semi-classical
differential scattering cross-section in a certain range of scattering
angles.  We briefly summarize in the next paragraph the geodesic
approximation for the low-energy classical dynamics of BPS $SU(2)$
magnetic monopoles.  Following that, we pursue the semi-classical
quantization of this system.

\smallskip
The low-energy classical dynamics of BPS $SU(2)$ magnetic monopoles can
be modeled using the geodesic approximation of N. Manton [6].  In this
scheme one makes use of the moduli space of static, finite-energy,
non-singular solutions $(A,\Phi)$ to the Bogomol'nyi equations on
${\bf R}^3$,
$${1\over 2}\epsilon_{ijk}F^A_{jk}=D^A_i\Phi.$$
The Lie algebra-valued gauge potential $A$ and Higgs field $\Phi$ are both
in the adjoint representation of $SU(2)$.
$D^A$ is the exterior covariant derivative
of the gauge potential, $A$, and $F^A$ is the gauge curvature.  Finite-energy,
non-singular solutions to the Bogomol'nyi equations are called BPS $SU(2)$
magnetic monopoles, and can be classified by a topological number interpreted
as the magnetic charge.  The geodesic approximation maintains that the
low-energy dynamics of two interacting magnetic monopoles is well-approximated
by the geodesic motion on the moduli space of all BPS magnetic monopoles
with a magnetic charge of two.  Fixing the centre-of-mass of the two
magnetic monopole system and factoring out by an irrelevant
overall phase, the moduli space $M^o_2$ is of real dimension four.
Atiyah and Hitchin made use of a natural, non-trivial $SO(3)$ action
on the moduli space $M^o_2$ to determine the unique, complete metric
structure on $M^o_2$ [7].  The metric can be put in the form
$$ds^2=f(r)^2\,dr^2+a(r)^2\sigma_1^2+b(r)^2\sigma_2^2
+c(r)^2\sigma_3^2.\eqno{(1)}$$
$\{\sigma_1,\sigma_2,\sigma_3\}$ is the usual dual basis for $so^*(3)$.
There is a certain amount of freedom in choosing $f(r)$, we follow
Atiyah and Hitchin and define $f(r)\equiv -abc$.
The equations of motion computed from (1) and the geodesic
approximation, which form the basis of our study, are:
$$\eqalign {{2bc\over f}{da\over dr}&= (b-c)^2-a^2\ \ \ \ (cyclic),\cr
{dM_1\over dt}&=\bigg({1\over b^2}-{1\over c^2}\bigg)M_2M_3
\ \ \ \ (cyclic),\cr
{d^2r\over dt^2}=-{1\over f}{df\over dr}\bigg({dr\over dt}\bigg)^2+&
{1\over f^2}\bigg({1\over a^3}{da\over dr}M_1^2+
{1\over b^3}{db\over dr}M_2^2+{1\over c^3}{dc\over dr}M_3^2\bigg).}
\eqno{(2)}$$
The first set of equations define the metric coefficients in (1), and
the remaining equations are the geodesic equations on the moduli space
$M^o_2$.  The work in [6,7] permits one to examine in detail the
classical soliton interaction dynamics of magnetic monopoles contained
within the prototype of the standard model: Yang-Mills-Higgs theory.
Note, however, that the representation is not that of the standard model.
In addition to its connections with particle physics, the geodesic
equations above form a non-integrable generalization [8] to the
familiar Euler-Poinsot equations for a rigid body---equations (2)
therefore should be of rather broad interest to physicists.

\smallskip
There are two totally geodesic (real) surfaces, $\Sigma_1$ and $\Sigma_{12}$,
in the moduli space, $M^o_2$, for which the scattering behaviour is
analytically well-understood [7].  These two cases contain the
`extremes' of the classical two-monopole dynamical behaviour, and as such
they are not expected to be the principle contributors to the differential
cross-section.
For $\Sigma_1$ all impact parameters, $\lambda$, lead to repulsive
scattering.  For $\lambda=0$, the scattering angle is at its maximum,
$\pi/2$, and, as the impact parameter increases the scattering angle
falls to zero.
The geodesic motion on $\Sigma_{12}$ is more complicated.
$\Sigma_{12}$ possesses scattering trajectories in three ranges (fig. 1).
The dynamics on $\Sigma_{12}$ are unusual in that the magnetic monopole
system can convert orbital angular momentum into electric charge
(angular momentum of the electromagnetic field), and back again.
In the first range (I, fig. 1), all orbital angular momentum is
permanently converted into electric charge, thereby scattering
at $\pi/2$ radians directly out of the scattering plane.
In range II, fig. 1, Atiyah and Hitchin define a new variable,
$\epsilon$, which is related to the impact parameter, $\lambda$, by
$\lambda=1+\epsilon$ [7], and compute the highest order term
for the scattering angle, $\theta$, as a function of $\epsilon$, given by
$$\theta(\epsilon)\sim\pi(2\epsilon)^{-{3\over 2}},
\eqno{(3)}$$
for $0<\epsilon\le\pi/2-1$.  In this range the monopoles are attractive.
We may write $\theta\equiv\theta_o+2\pi j\sim\pi(\epsilon_j/2)^{-3/2}$
where $0<\theta_o\le\pi$, $j\in{\bf Z}^+\cup\{0\}$, and solve for
$\epsilon_j$ (real):
$$\epsilon_j\sim {1\over 2}\bigg({\pi\over \theta_o+2\pi j}
\bigg)^{2\over 3}.\eqno{(4)}$$
Finally, in the third range of impact parameters (III, fig. 1), the
scattering angle for all impact parameters is less than or equal to
$\theta=\pi/3$.  In IIIa the scattering angle increases repulsively,
with impact parameter, from zero to $\theta=\pi/3$.  In IIIb, the
scattering angle decreases monotonically to zero with increasing
impact parameter.  A rainbow therefore exists at $\pi/3$ and also
at $2\pi/3$ (from the particle reflected through the centre-of-mass).
The impact parameter associated with the rainbow defines the
demarcation between IIIa and IIIb.  The range of scattering angles
for which we compute the differential cross-section will lie between
the rainbows at $\pi/3$ and $2\pi/3$.

\smallskip
Assuming that $\theta_o$ is away from classical rainbows,
the semi-classical scattering amplitude is [10]
$$f(\theta_o)=\sum_{j=0}\sqrt{c_j}e^{i(S_j/\hbar-\pi\mu_j/2)}.
\eqno{(5)}$$
The sum runs over all classical scattering trajectories, $\Gamma_j$,
scattering into $\theta_o$.  From (3) we see that on $\Sigma_{12}$ there
are an infinite number of classical trajectories with scattering angle
$\theta_o$ for BPS $SU(2)$ monopole scattering.  Of course, depending
on $\theta_o$, there may be other contributing paths not on $\Sigma_12$.
Let us consider for a moment, however, only those scattering trajectories
$\Gamma_j$ on $\Sigma_{12}$.
For the range of angles we study in this paper, $\pi/3<\theta<\pi/2$
and $\pi/2<\theta<2\pi/3$, impact parameters in region III do not contribute
to the differential cross-section.  Likewise, impact parameters in region I
contribute only to $\theta=\pi/2$.  Since the semi-classical scattering
amplitude (5) breaks down at $\theta=\pi/2$, other methods must be used for
this scattering angle (see [11], for example).  Therefore on $\Sigma_{12}$
only impact parameters in region II contribute to the scattering angles
$\pi/3<\theta<\pi/2$ and $\pi/2<\theta<2\pi/3$.
The Maslov index, $\mu_j$, in equation (5) counts the number of caustics
in $\Gamma_j$; $\mu_j=0$ for all trajectories on $M^o_2$.
The Action $S_j$ in (5) is given by $-\int_{\Gamma_j}{\bf p}.\,d{\bf q}$.
Finally, $c_j$ is the contribution to the scattering amplitude of the
trajectory $\Gamma_j$ with $\epsilon=\epsilon_j$ where $c_j$ is given by
$$c_j\equiv\bigg\vert{d\theta\over d\epsilon}\bigg\vert^{-1}(\epsilon_j)
\sim{1\over 3\pi}(2\epsilon_j)^{5\over 2}={1\over 3\pi}
\bigg({\pi\over\theta_o+2\pi j}\bigg)^{5\over 3},\eqno{(6)}$$
using (3) and (4).

\smallskip
Numerical simulation in [1] involving all the classical paths of the
Atiyah-Hitchin equations in (2), that is, not just the trajectories on
$\Sigma_{12}$, suggests that there are no classical rainbows round
$\theta_o=\pi/2$, and therefore we can view (5) as being valid round
$\pi/2$.  The known rainbow at $\theta_o=\pi/3$ was successfully
identified by the same numerical procedure.  Further numerical
simulation found that the classical and quantum magnetic monopole
differential cross-section were in excellent agreement at
$\theta_o=\pi/2$ in the centre-of-mass frame.  The numerical
results discount Mott interference effects [1].
We use this numerical result to write the differential cross-section
at $\theta_o$ as
$${d\sigma\over d\Omega}(\theta_o)=\vert f(\theta_o)\vert^2
+\vert f(\pi-\theta_o)\vert^2.\eqno{(7)}$$
We refer the justifiably sceptical reader to [1] for a detailed
discussion of the error analysis, extensive checks on the numerical
code, and a theoretical argument tying the result to the lack of
singularities on the configuration space (moduli space).
We note that singularities are a necessary feature of point-particle
configuration spaces, but not of solitonic configuration spaces.
Therefore we believe that similar topological arguments can be
expected for many other solitons, in particular those in [2-5].
Using equation (5) and $\mu_i=0$, the norm of the scattering amplitude
becomes
$$\vert f\vert^2\approx\sum_{j=0}^\infty c_j+
2{\rm Re}\sum_{i<j}^\infty\sqrt{c_ic_j}\exp{i(S_i-S_j)/\hbar}.
\eqno{(8)}$$
A detailed understanding of the Action difference in the phase
is out of our reach, but there is much that can still be inferred
about the semi-classical cross-section.

\smallskip
The Riemann-Lebesgue Lemma, equation (6), and equation (7) imply that
in the classical limit ($\hbar\to 0$) the differential cross-section
restricted to $\Sigma_{12}$ is
$$\eqalign {\sum_{j=0}^\infty\bigg( c_j(\theta_o)+c_j(\pi-\theta_o)\bigg)
&\sim{1\over 3\pi}\sum_{j=0}^\infty\bigg[\bigg(
{\pi\over\theta_o+2\pi j}\bigg)^{{5\over 3}}+\bigg(
{\pi\over(\pi-\theta_o)+2\pi j}\bigg)^{{5\over 3}}\bigg]\cr
&={1\over 3\pi}\bigg[\zeta\bigg({5\over 3},{\theta_o\over 2\pi}\bigg)
+\zeta\bigg({5\over 3},{\pi-\theta_o\over 2\pi}\bigg)\bigg].}\eqno{(9)}$$
We have rewritten the series using the generalized Riemann
$\zeta$-function, $\zeta(s,a)$ [12].
The classical differential cross-section as approximated in (9) is
well-defined for all $\theta_o$ away from classical rainbows.  However,
it is not difficult to see that the second term in (8), the cross-term,
is {\it not} absolutely convergent.
The semi-classical scattering amplitude is only conditionally
convergent since the absolute double series can be shown to be
equal to
$$2\sum_{i<j}\vert\sqrt{c_ic_j}\vert(\theta_0)\sim{1\over 3\pi}
\bigg[\zeta^2\bigg({5\over 6},{\theta_o\over
2\pi}\bigg)-\zeta\bigg({5\over 3},{\theta_o\over 2\pi}\bigg)\bigg].
\eqno{(10)}$$
This expression diverges with $\zeta(5/6,\theta_o/2\pi)$ for
all values of $\theta_o$.  The scattering amplitude (5) is
also only conditionally convergent---the absolute series
is a divergent hyperharmonic series for all $\theta_o$.
The conditionally convergent differential cross-section therefore
depends on the order in which contributions from the classical
paths are summed.
Contributions to the differential cross-section from trajectories away
from $\Sigma_{12}$ will not alter the conditional convergence.
Thus although the magnetic monopoles and their dynamics are both
non-singular, the semi-classical scattering cross-section still
retains the mildest form of divergence: conditional convergence.  We
remove any remaining difficulties with the convergence of the theoretical
cross-section using $\zeta$-function regularization; the series is
made absolutely convergent by analytically continuing the
$\zeta$-function into the complex plane.  If we now maximize the
quantum effects by introducing a WKB condition on the scattering
trajectories, $\Delta S=2\pi n\hbar$, the semi-classical differential
cross-section due to scattering initial conditions on $\Sigma_{12}$ is
then approximated by
$${d\sigma\over d\Omega}\bigg\vert_{\Sigma_{12}}\approx
{1\over 3\pi}\bigg[\zeta^2\bigg({5\over 6},{\theta_o\over 2\pi}\bigg)
+\zeta^2\bigg({5\over 6},{\pi-\theta_o\over 2\pi}\bigg)\bigg].
\eqno{(11)}$$
The acceptability of this assumption is reflected in our final results.

\smallskip
The contributions to the differential cross-section when we consider
all scattering data are dominated by the scattering between the geodesic
subsurfaces $\Sigma_1$ and $\Sigma_{12}$.
The dynamical behaviour between $\Sigma_1$ and $\Sigma_{12}$ is studied
numerically in [8].  The following observations can be made
from that numerical study:
first, some of the dynamics in range I, fig. 1, rotate into
the $\pi/2$ scattering on $\Sigma_1$; second, the dynamics between
$\Sigma_1$ and $\Sigma_{12}$ change smoothly and qualitatively
look much like the dynamics on $\Sigma_{12}$.  To get to $\Sigma_1$
from $\Sigma_{12}$, simulation suggests that the impact parameter
ranges I, II, and IIIa must collapse smoothly onto zero impact
parameter while some of the scattering in region I, fig. 1, rotates
into large angle scattering on $\Sigma_1$.
Therefore, the critical impact parameters, $\lambda_1$,
$\lambda_2$, and $\lambda_3$ which separate the scattering ranges must
vanish as the dynamics approaches that on $\Sigma_1$.  This has the effect
of truncating classical paths in II.  The truncation of classical paths
can be managed if we assume that the highest order term in
$\theta(\epsilon)$ (see equation (3)) is of the same order as on
$\Sigma_{12}$.  In those circumstances,
$$\sum_{j=N}^\infty c_j\sim\sum_{j=N}^\infty{1\over 3\pi}\bigg(
{\pi\over \theta_o+2\pi j}\bigg)^{5\over 3}={1\over 3\pi}
\zeta\bigg({5\over 3},{\theta_o+2\pi N\over 2\pi}\bigg),$$
and the cross-term in (10) can be similarly rewritten.  We make the
Riemann zeta-function periodic in the second argument in $\zeta(s,a)$,
for otherwise the cross-section diverges for all $\theta_o$.  The
periodicity $\Delta a=1$ is suggested using results from a
partial wave analysis due to Schroers [12]: the magnetic monopole
quantum differential cross-section is expected to be very close
to the integrable quantum Taub-NUT differential cross-section at
$\theta_o=\pi/2$ (CoM frame).  The quantum Taub-NUT cross-section
can be computed explicitly by virtue of its integrability to give 4.0
at $\theta_o=\pi/2$ [10].  With periodicity $\Delta a=1$, the
differential scattering cross-section is constant for
all values of an internal relative phase, $0\le\psi<2\pi$, which
parametrises the scattering between $\Sigma_1$ and $\Sigma_{12}$ [8].
Therefore the total differential cross-section for solitonic
magnetic monopoles is given by $\pi$ (not $2\pi$, which double counts
the particles) times the differential cross-section in (11):
$${d\sigma\over d\Omega}(\theta_o)\approx{1\over 3}\bigg[
\zeta^2\bigg({5\over 6},{\theta_o\over 2\pi}\bigg)+
\zeta^2\bigg({5\over 6},{\pi-\theta_o\over 2\pi}\bigg)\bigg].
\eqno{(12)}$$
Round $\theta_o=\pi/2$ the approximation to the semi-classical
cross-section (12) gives approximately 4.48, which compares
favourably with 4.0 from the exact quantum Taub-NUT cross-section.

\smallskip
We plot in Figure 2 the semi-classical differential cross-section (12)
(dark curve), the classical limit (9) (grey curve), and the
solitonic differential cross-section computed numerically on a
Connection Machine in [1] (data points).  Only
a finite number of monopoles (32,000) is used to construct
the classical solitonic cross-section.  Increasing the number of
monopoles allows us to include monopole scattering events with
larger impact parameters.  Small angle events will be more
numerous, pushing the sides of the cross-section up and
simultaneously lowering the probability of large angle scattering
events.  The classical solitonic cross-section between $\pi/3$ and
$2\pi/3$, therefore, drops visibly towards the semi-classical
cross-section, and away from the classical limit cross-section
as the number of particles increases.  Taking this into account,
the approximation to the semi-classical differential cross-section
between the rainbows at $\pi/3$ and $2\pi/3$ is in good
agreement with the classical solitonic cross-section,
and in conflict with the classical limit.  The convergence of
the classical solitonic and semi-classical cross-sections at
$\theta_o=\pi/2$ is discussed in [1].  We note the
distinctive (pseudo-) quantum hump effect at $\theta=\pi/2$ in the
classical solitonic differential cross-section, which is seen in
the semi-classical cross-section as well (despite the lack of Mott
interference terms in the cross-section (7)).  We comment also on the
WKB assumption in equation (11) which maximized the quantum effects in
the differential cross-section.  We now see that reducing the quantum
effects would push the semi-classical cross-section up, away from its
expected value of 4.0 at $\theta_o=\pi/2$.  This is our justification
for introducing WKB conditions onto scattering trajectories.
Finally, we conclude by saying that both the classical scattering
behaviour here and the classical dyonium bound states with discrete
energies in [1] suggest some truth in Einstein's conjecture that
quantum dynamics should emerge from a highly constrained classical
theory.

\vskip 0.5in
\openup -2\jot
\noindent
\centerline {\bf References}
\medskip
\item{[1]} M. Temple-Raston, D. Alexander, {\it Nucl. Phys.}
{\bf B397}, 195 (1993).
\smallskip
\item{[2]} P.J. Ruback, {\it Nucl. Phys.} {\bf B334}, 669 (1988);
T.M. Samols, {\it Comm. Math. Phys.} {\bf 145}, 149 (1992).
\smallskip
\item{[3]} N. Manton, {\it Phys. Lett.} B {\bf 192}, 177 (1987);
P.M. Sutcliffe, {\it Nonlinearity} {\bf 4},
1109 (1991).
\smallskip
\item{[4]} R.A. Leese, {\it Nucl. Phys.} {\bf B334}, 33 (1990);
W.J. Zakrzewski, {\it Nonlinearity} {\bf 4}, 429 (1991).
\smallskip
\item{[5]}  C. Rosenzweig, A.M. Srivastava, {\it Phys. Rev. D}
{\bf 43}, 4029 (1991).
\smallskip
\item{[6]} N.S. Manton, {\it Phys. Lett.} B {\bf 110}, 54 (1982).
\smallskip
\item{[7]} M.F. Atiyah, N.S. Hitchin, The Geometry and Dynamics of
Magnetic Monopoles (Princeton U.P., Princeton, 1989).
\smallskip
\item{[8]} M. Temple-Raston, {\it Phys. Lett.} B {\bf 206},
503 (1988).
\smallskip
\item{[9]} M. Temple-Raston, {\it Phys. Lett.} B {\bf 213},
168 (1988); M. Temple-Raston, {\it Nucl. Phys.} {\bf B313},
447 (1989).
\smallskip
\item{[10]} G. Gibbons, N. Manton, {\it Nucl. Phys.} {\bf B274}
(1986) 183.
\smallskip
\item{[11]} M.C. Gutzwiller, Chaos in Classical and Quantum Mechanics
(Springer-Verlag, New York, 1990).
\smallskip
\item{[12]} B.J. Schroers, {\it Nucl. Phys. B} {\bf 367} (1991) 177.
\smallskip
\item{[13]} E.T. Whittaker, G.N. Watson, A course of modern analysis,
4th ed. (Cambridge U.P., Cambridge, 1952).
\bigskip
\centerline{\bf Figure Captions}
\medskip
\noindent
{\bf Figure 1.}  Ranges of impact parameters where the scattering
dynamics changes significant in character.
\smallskip
\noindent
{\bf Figure 2.}  Magnetic monopole differential cross-sections (DCS).
The solid line is the semi-classical DCS, the grey line is the classical
limit of the semi-classical DCS, and the data points give the classical
solitonic DCS obtained numerically with 32,000 monopoles.